\documentclass[showpacs,twocolumn,longbibliography,amsmath,amssymb,superscriptaddress,pra]{revtex4-1}

\usepackage{color}
\usepackage{graphicx}
\usepackage{amssymb}
\usepackage{amsmath}
\usepackage{mathrsfs}
\newcommand{\ud}{\mathrm{d}}

\newcommand{\pdern}[3]{\frac{\partial^#3#1}{\partial#2^#3}}
\newcommand{\pder}[2]{\frac{\partial#1}{\partial#2}}

\usepackage{hyperref}
\graphicspath{ {./figures/} {./}}

\begin{document}

\title{Recurrence in the high-order nonlinear Schr{\"o}dinger equation: a low dimensional analysis}

\author{Andrea Armaroli}
\email{andrea.armaroli@unige.ch}
\affiliation{GAP-Nonlinear, Universit{\'e} de Gen{\`e}ve, Chemin de Pinchat 22, 1227 Carouge, Switzerland}
\affiliation{ISE, Universit{\'e} de Gen{\`e}ve, Boulevard Carl-Vogt 66, 1205 Gen{\`e}ve, Switzerland}
\author{Maura Brunetti}
\affiliation{GAP-Nonlinear, Universit{\'e} de Gen{\`e}ve, Chemin de Pinchat 22, 1227 Carouge, Switzerland}
\affiliation{ISE, Universit{\'e} de Gen{\`e}ve, Boulevard Carl-Vogt 66, 1205 Gen{\`e}ve, Switzerland}
\author{J{\'e}r{\^o}me Kasparian}
\affiliation{GAP-Nonlinear, Universit{\'e} de Gen{\`e}ve, Chemin de Pinchat 22, 1227 Carouge, Switzerland}
\affiliation{ISE, Universit{\'e} de Gen{\`e}ve, Boulevard Carl-Vogt 66, 1205 Gen{\`e}ve, Switzerland}
\date{\today}

\begin{abstract}
We study a three-wave truncation of the high-order nonlinear Schr{\"o}dinger equation for deep-water waves (HONLS, also named Dysthe equation). We validate the model by comparing it to numerical simulation, we  distinguish the impact of the different fourth-order terms and classify the solutions according to their topology. This allows us to properly define  the temporary spectral \emph{upshift} occurring in the nonlinear stage of Benjamin-Feir instability and  provides a tool for studying further generalizations of this model.

\end{abstract}
\maketitle

\section{Introduction}

The nonlinear Schr{\"o}dinger equation (NLS) is a universal model which applies to deep-water waves, nonlinear optics, plasma physics and Bose-Einstein condensates among others \cite{Osborne2009,SulemNLSBook}. It is an integrable partial differential equation which gives us access to a set of powerful mathematical techniques \cite{Zakharov1972,Biondini2016}. 
Particularly, in hydrodynamics, where the wave steepness plays the role of perturbation parameter, {the NLS is an equation for the envelope of a slowly modulated carrier wave. It is derived as the  compatibility condition at the third-order in the multiple-scale expansion of the Euler equations for an incompressible inviscid fluid with uniform density}. In order to overcome its intrinsic narrow-band nature, nearly 40 years ago a fourth-order generalization of the NLS was proposed in \cite{Dysthe1979,Janssen1983}. {In the following, we consider only the one-dimensional propagation, as in \cite{Lo1985,Kit2002,Zhang2014}, and  refer to this model as high-order NLS (HONLS), which was shown to accurately describe the experimental results collected in water tanks \cite{shemer2002,shemer2010,Chabchoub2013}}. The main drive to develop such a model was to better reproduce the properties of the the ubiquitous Benjamin-Feir instability (BFI) \cite{Benjamin1967,Zakharov2009}, i.e.~the well-known growth of oscillations on top of a uniform Stokes wave. 
{For example}, the spectral downshift observed in BFI \cite{Lake1977} can be often explained in the HONLS framework.
However, in the last decades much effort has been devoted to discuss the properties of this model and to include other effects, such as viscosity or wind  \cite{Miles1957,Dias2008,Touboul2010,Onorato2012,Brunetti2014,Brunetti2014b,Schober2015,Carter2016,Toffoli2017,Eeltink2017}. 

As far as the mathematical properties of the HONLS are concerned, we recall that the conventional reference frame change, which allows one to write a time-like HONLS (where the evolution variable is space), leads to two different forms: the equation describing the evolution of the envelope of the surface elevation (thereafter denoted HONLSe) and the one pertaining to the envelope of the velocity potential (HONLSp). The former possesses less conservation laws than the latter: this is a consequence of the choice of non-canonical variables.

Canonical variables involve both the surface elevation and the velocity  potential \cite{Zakharov1968,Stiassnie1984a,Gramstad2011,Fedele2012}, but only the former is experimentally accessible. It is thus interesting to understand what broken integrability implies on directly measurable quantities.
A hint of an underlying integrable evolution is that the HONLS exhibits a nearly perfect recurrence of the initial state. Moreover during this cyclic behavior the spectral mean of the surface elevation is temporarily upshifted while the spectral peak is downshifted \cite{Lake1977,Lo1985,Tulin1999}.

In this work we will study the behavior of recurrence in the  two forms of HONLS and characterize the origin of the temporary peak downshift. After having recalled in section \ref{sec:HONLS} the main properties of the model, we present (section \ref{sec:CMT}) a three-wave truncation \cite[p. 527]{WhithamBook} \cite{Trillo1991c}, by means of which we find a closed form for the equations which rule the nonlinear behavior of BFI, obtain an approximate one-degree-of-freedom phase space, where the evolution of experimental quantities can be mapped onto, and clarify what a sound definition of spectral shift should be. 

This approach proved effective in other nearly-integrable systems \cite{Conforti2016}, where exact solutions cannot be constructed. This is the case of HONLS: the Akhmediev breather \citep{Akhmediev1987b,Chabchoub2014},  the Peregrine soliton \cite{Peregrine1983,Kibler2010}, conventionally regarded as prototypes of rogue waves \cite{Akhmediev2009,Dudley2014}, only represent approximate solutions \cite{Zakharov2010}. The non-integrability appears nevertheless to be an important ingredient for the appearance of extreme events (e.g.~in optical fibers \cite{Solli2007,Armaroli2015}).

\section{High-order nonlinear Sch{\"o}dinger equations}
\label{sec:HONLS}
\subsection{Model equations}
{We consider the propagation of water waves along the direction $x$  in a one dimensional tank, the width of which is supposed small in order not to observe two-dimensional structures on the water surface. The water depth appears only in the coefficients of dispersion and nonlinearity and completely disappears in the limit of an infinitely deep tank. In an incompressible fluid the nonlinearity is introduced by boudary conditions on the free surface and the bottom of the tank (assumed perfectly rigid). Only the free-surface motion is important to describe the wave propagation. The normalized time-like HONLS can be derived from Eqs.~(4.1-4.3) of \cite{Dysthe1979} following the approach of  \cite{Janssen1983,Lo1985,Kit2002,Zhang2014}, i.e.~by changing the reference frame, neglecting derivatives in the transverse variable, and eliminating the third-order dispersion by replacing the expression of the NLS. It reads as}
\begin{equation}
	\pder{a}{\xi} + i \frac{1}{2} \pdern{a}{\tau}{2} + i|a|^2a ={\epsilon} \left\{f[a]+\beta g[a]+h[a]\right\},	
	\label{eq:Dysthe1}
	\end{equation}
where the adimensional quantities are obtained from dimensional ones as $a = A/A_0$, $\tau=t/T_0$, $\xi=x/L_0$. $A$ is the complex envelope of surface elevation, $t$ and $x$ are time and propagation distance in a frame moving at the group velocity of the carrier wave. The normalization constant $A_0$ is chosen by setting a value of the steepness $\epsilon=A_0k_0/\sqrt{2}$, $k_0$ being the carrier wave-number. Finally $T_0 = 1/(\omega_0\epsilon)$, $L_0 = 1/(2\epsilon^{2}k_0)$ with $\omega_0=\sqrt{g k_0}$ the angular frequency of the wave in the limit of infinitely deep water.  Thanks to those definitions, $a=\mathcal{O}(1)$ and $\epsilon$ controls the relative importance of higher-order corrections. {It  thus makes explicit the hierarchy of orders in the perturbation expansion of Euler equations: the NLS comes from third-order terms in $\epsilon$ and HONLS is thus fourth-order.}  We define $f[a]\equiv 8|a|^2\pder{a}{\tau}$, $g[a] \equiv 2 a^2
	\pder{a^*}{\tau}$ and $h[a]\equiv 2i	a \mathcal{H}\left[\pder{|a|^2}{\tau}\right]$ ($\mathcal{H}[\cdot]$ denotes the Hilbert transform). $\beta$ is used to distinguish the HONLSe ($\beta=1$) from the HONLSp ($\beta=0$)\cite{Lo1985}. 
{We reconstruct, from the HONLSe, the surface elevation $\eta(x,t) = \mathrm{Re}\,\{a(x,t)\exp\left[i\left(k_0 x - \omega_0 t\right)\right)\}$	 and, from the HONLSp, the velocity potential at the free surface $\bar\phi(x,z=\eta(x,t),t) =  \mathrm{Re}\,\{a(x,t)\exp\left[i\left(k_0 x - \omega_0 t\right)\right)\}$.} 
This latter conserves not only the norm $N \equiv\int_{-\infty}^{\infty}{|a|^2\ud \tau}$ (the only conserved quantity of HONLSe), but also the momentum  $P \equiv\frac {i}{2}\int_{-\infty}^{\infty}{(a^*_\tau a-a_\tau a^*)\ud \tau}$ and the  Hamiltonian 	
	$E \equiv E_0-\epsilon E_1$, with $E_0=\frac{1}{2}\int_{-\infty}^{\infty}{\left(|a_\tau|^2- |a|^4\right) \ud\tau}$ and $E_1=i\int_{-\infty}^{\infty}{|a|^2\left[2(a_\tau^* a - a_\tau a^*) - \mathcal{H}[|a|^2]_\tau\right] \ud\tau}$. 

\subsection{Numerical analysis}
We find it useful to briefly recall the properties of the full nonlinear evolution of BFI after the initial exponential growth phase.
Since the excitation of a harmonically perturbed Stokes wave represents a good approximation of the Akhmediev breather, we solve the HONLSe with initial condition $a(0,\tau) = (1-\sqrt{\tilde \eta})+\sqrt{2\tilde\eta}\cos(\Omega\tau+\tilde\phi)$, with $\tilde\eta=1\times 10^{-2}$, $\tilde\phi=\pi/4$ and $\Omega=\Omega_\mathrm{M}=\sqrt{2}(1-\sqrt{2}\frac{3}{2}\epsilon)$, i.e.~the frequency of maximum instability growth, which depends only on  $h[a]$   \cite{Dysthe1979} (see also the derivation below).  We use $\epsilon=\sqrt{2}/20$ as initial steepness.
We numerically solve Eq.~\eqref{eq:Dysthe1} by means of the 3rd-order Runge-Kutta (RK) scheme embedding the conventional 4th-order RK applied in the interaction picture formulation of the HONLSe   \cite{Balac2013}.  
\begin{figure}[htpb]
\centering
\includegraphics[width=.5\textwidth]{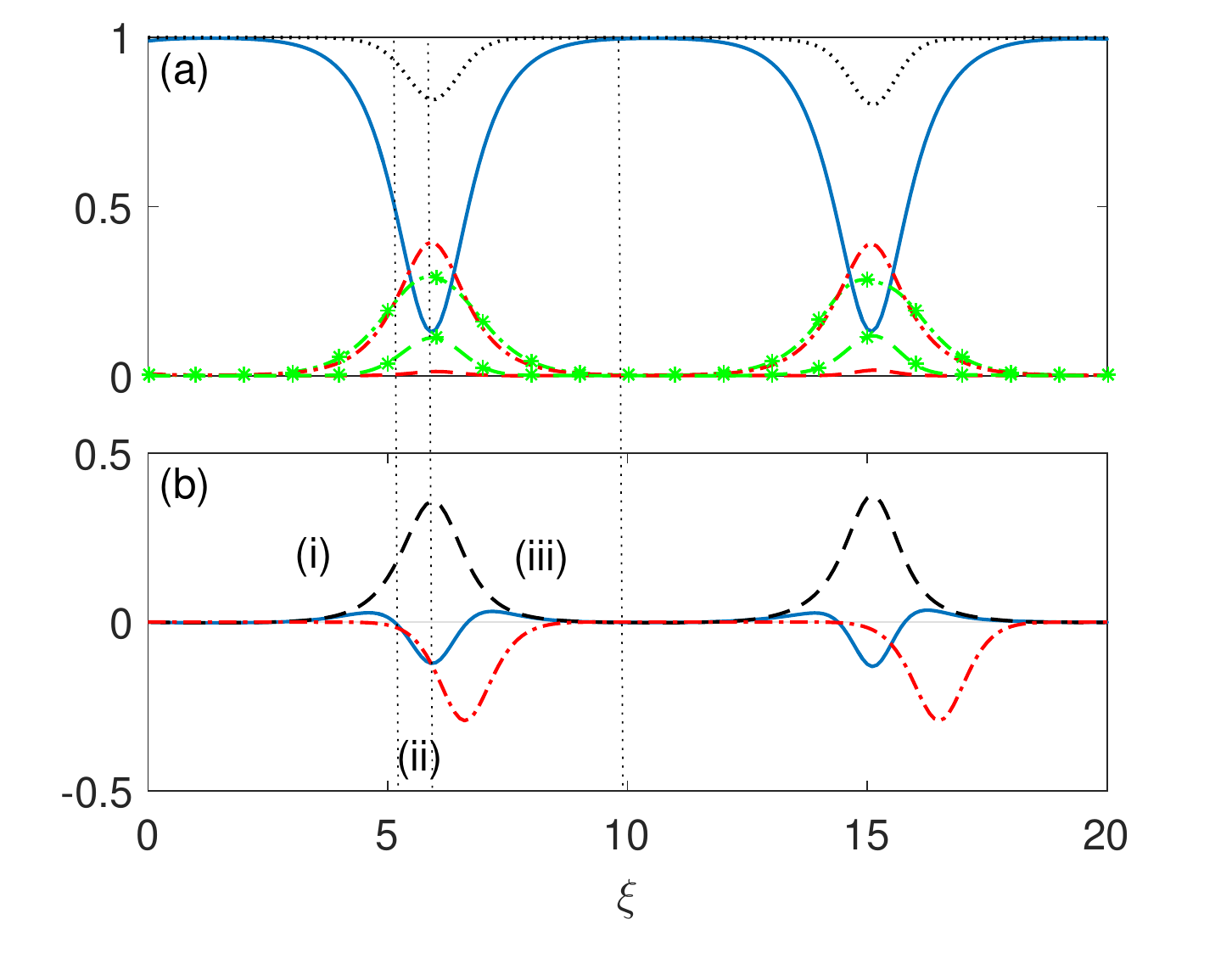}
\caption{(a) Numerical results of HONLSe simulations. Evolution of the squared amplitude of the Stokes wave (blue solid line), its unstable sideband pair  ($\pm\Omega$, dash-dotted lines) and second order sidebands ($\pm2\Omega$, dashed lines). Red (green with asterisks) lines represent the low (high) frequency sideband. The black dotted line represent the sum of square amplitudes of Stokes wave and its unstable sidebands. (b) The corresponding  relative unbalance of sideband amplitudes $\Pi$ (solid blue for HONLSe, dash-dotted red in the  HONLSp) and spectral shift $P/N$ (dashed black, full HONLSe; not reported for HONLSp). 
The labels (i-iii) correspond to three stages of  recurrence in HONLSe mentioned in the main text. 
}
\label{fig:dysthe1}
\end{figure}
In Fig.~\ref{fig:dysthe1}(a) we show the evolution of the Stokes wave and its lateral sidebands of first (the unstable modes, at $\pm\Omega$) and second order ($\pm2\Omega$ generated by the four-wave mixing process $\omega_0\pm\Omega + \omega_0\pm\Omega-\omega_0 \to \omega_0\pm2\Omega$). The recurrence is almost perfect and can be separated in three phases: (i) the unstable modes grow with a predominant upshift, see the dash-dotted lines in panel (a) and also the blue line in Fig.~\ref{fig:dysthe1}(b), which corresponds to the sideband unbalance $\Pi\equiv\Omega(|\hat a(\Omega)|^2-|\hat a(-\Omega)|^2)$ ($\hat a$ is the Fourier-transform of $a$);  (ii) the  second order sidebands  start growing due to four-wave mixing; this effect is unbalanced towards $2\Omega$ (so that the spectral mean grows) and  contribute to the opposite-side unstable mode ($-\Omega$) by four-wave mixing, thus we observe a peak downshift as soon as the Stokes wave is overcome by the unstable modes; (iii) at maximum conversion the process reverses to the initial perturbed plane wave.

In Fig.~\ref{fig:dysthe1}(b) we also plot the ratio of  momentum and norm $P/N$.  We can derive an integral expression for the evolution of  this quantity as:
\begin{equation}
	\frac{\ud}{\ud \xi}\frac{P}{N}= -4\beta\epsilon\frac{R}{N}
	\label{eq:PoverN}
\end{equation}
with $R\equiv \frac{i}{2}\int_{-\infty}^{\infty}{|a|^2(a^*_{\tau\tau} a - a_{\tau\tau}a^*)\ud \tau}$. 

If $\Pi$ changes sign and unstable sidebands dominate over the Stokes wave, we can state that the spectral peak has shifted, so this variable describes what is the experimental peak frequency. Instead $P/N$ is, by definition, an average over all stable and unstable modes and represents the spectral mean: in Fig.~\ref{fig:dysthe1}(b) we observe its temporary passage in the positive region for the HONLSe (black dashed line). 

$P$ is conserved for $\beta=0$: it is thus apparent that $g[a]$ is associated to the temporary growth of $P/N$. The spectral peak downshift is instead observed in both models (compare blue solid and red dash-dotted lines in Fig.~\ref{fig:dysthe1}(b)), thus it can be associated to $f[a]$, which appears  as an intensity-dependent increase of group velocity  \cite{Lo1985,Goullet2011}. 
As it was shown in  \cite{Akylas1989,Zhang2014}, this term induces an almost homogeneous drift in the recurrence. 
Given the dispersion relation, an increase in group velocity $C_g=g/(2\omega_0)$ is equivalent to a decrease in the instantaneous frequency. Beyond this qualitative explanation, a numerical analysis of the solutions of HONLSp allows us to fit, close to $\Omega_\mathrm{M}$, the maximum shift with $\Delta\Omega_\mathrm{max}=-3\epsilon(1-2(\Omega-\Omega_\mathrm{M}))$: for a single unstable mode, less conversion to sidebands implies a smaller downshift. We will refer to the effect of $f[a]$ as kinematic.

Notice that the recurrence  period exhibits (see Fig.~\ref{fig:dysthe1}(b)) a ($\simeq 10\%$) walk-off between HONLSe and HONLSp: here HONLSp lags behind HONLSe, but a different choice of the initial relative phase $\tilde\phi$ leads to either positive or negative differences. 
We also observe  (Fig.~\ref{fig:dysthe1}(a)) that  only a moderate proportion of the energy is converted to higher-order sidebands (about $20\%$).

\begin{figure}
\centering
\includegraphics[width=.5\textwidth]{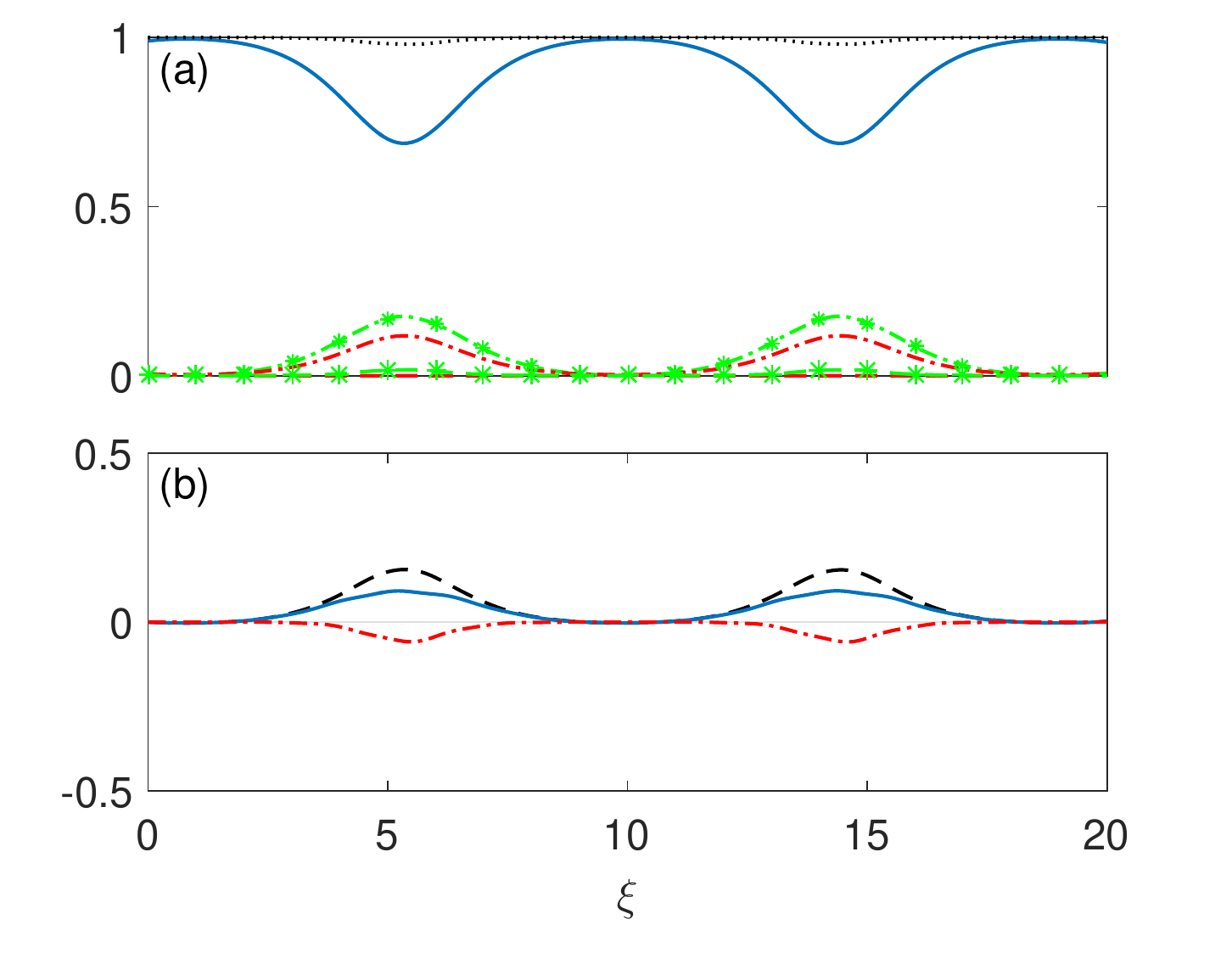}
\caption{Same as Fig.~\ref{fig:dysthe1}, but with $\Omega=1.6$.  Notice that the growth of unstable modes is much smaller and the kinematic peak downshift (still noticeable in HONLSp) cannot compensate the mean upshift in HONLSe. }
\label{fig:dysthe2}
\end{figure}
Finally, we repeat the numerical experiment for a unstable frequency closer to the instability margin, $\Omega=1.6$, where the BFI growth rate is smaller (see below). The relative impact of unstable modes is reduced and the upshift for $\beta=1$ dominates the always present downshift, see Fig.~\ref{fig:dysthe2}: notice that the blue line in panel (b) does not cross into negative values anymore.



\section{Three wave truncation}
\label{sec:CMT}
Given the limited conversion to higher-order sidebands, we conjecture that the dynamics of Eq.~\eqref{eq:Dysthe1} can be captured by a dynamical system with a finite number of degrees of freedom. 
We assume a limited tank length or, equivalently, a careful temporal profile of the initial condition in order to prevent that the unavoidable mixing products (at $\pm n\Omega$, with integer $n$) generated during propagation fall inside the instability range.
This single unstable mode regime corresponds to  the following Ansatz
\begin{equation}
a(\xi,\tau) = A_0(\xi) + A_1(\xi)e^{-i \Omega \tau} + A_{-1}(x)e^{i \Omega \tau}
\label{eq:3wAnsatz}
\end{equation}
where $A_0$ thus describes the evolution of the Stokes wave complex amplitude, while $A_{\pm 1}$ are the amplitudes of the unstable sidebands.
\begin{figure}[hptb]
\centering
\includegraphics[width=0.5\textwidth]{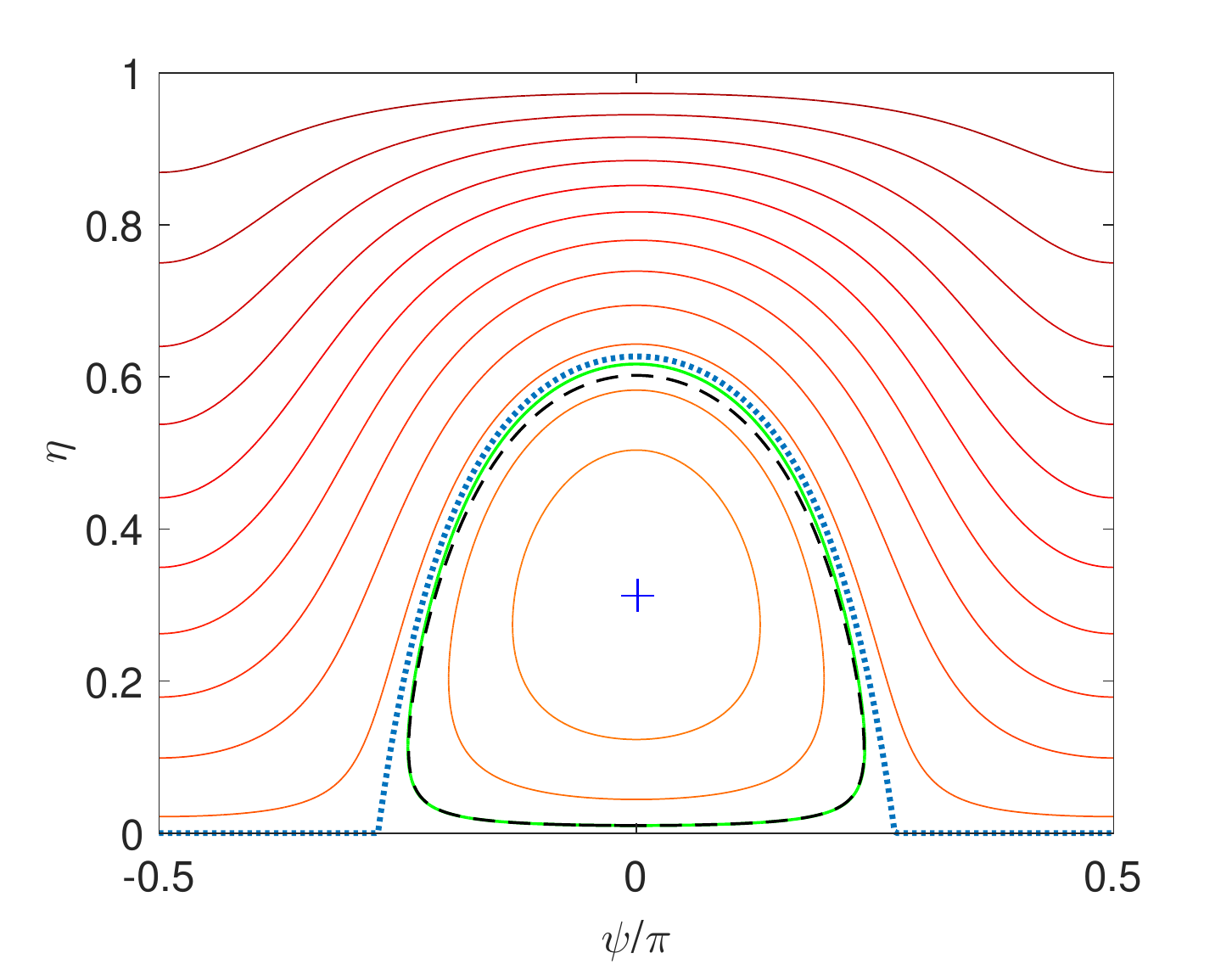}
\caption{Representation of the curves of constant $H$ (level sets, Eq.~\eqref{eq:hamiltonian}) in phase plane $(\psi,\eta)$. The dotted blue line represent the separatrix ($H=0$), the blue cross identifies the center fixed point; the level set corresponding to $\eta(0)=0.01$, $\psi(0)=0$, $\epsilon=\sqrt{2}/20$ is shown in green, while the dashed black line represents the actual solution of Eq.~\eqref{eq:1Dnonintegrable}. Notice that the superposition of the curves is almost perfect everywhere. }
\label{fig:3wavephasespace}
\end{figure}
By replacing the Ansatz of Eq.~\eqref{eq:3wAnsatz} inside Eq.~\eqref{eq:Dysthe1} we obtain a system of three complex equations 
\begin{widetext}
\begin{equation}
\begin{aligned}
	\dot{A}_0 &= -i\left(\left|A_0\right|^2+2\left|A_1\right|^2+2\left|A_{-1}\right|^2\right)A_0 - 2iA_0^*A_1A_{-1}  \\
	&+ i{\epsilon\Omega}\left\{-8(|A_1|^2-|A_{-1}|^2)A_0 + 4\beta(|A_1|^2-|A_{-1}|^2)A_0 
	+2\, s \left[(|A_{-1}|^2 + |A_{1}|^2)A_0 + 2A_0^*A_{-1}A_1\right]\right\}
\\
		\dot{A}_1 &= i\frac{\Omega^2}{2}A_1 - 
		i\left(\left|A_1\right|^2+2\left|A_0\right|^2
		+2\left|A_{-1}\right|^2\right)A_1-iA_0^2A_{-1}^*
		\\
		&+ i{\epsilon\Omega}
		\left\{-8(|A_0|^2+|A_{1}|^2)A_1 + 2\beta(|A_1|^2-2|A_{-1}|^2)A_1
		- 2\beta A_0^2 A_{-1}^*+
		2\,  s\left[\left(|A_0|^2 + 2 |A_{-1}|^2\right) A_1 + A_0^2A_{-1}^*\right] 
	\right\}	
	\\			
		\dot{A}_{-1} &= i\frac{\Omega^2}{2}A_{-1} - 
		i\left(\left|A_{-1}\right|^2+2\left|A_0\right|^2
		+2\left|A_1\right|^2\right)A_{-1}- iA_0^2A_{1}^*
		 \\
		&+ i{\epsilon\Omega}
		\left\{8(|A_0|^2+|A_{-1}|^2)A_{-1} - 2\beta(|A_{-1}|^2-2|A_{1}|^2)A_{-1} + 2 \beta A_0^2A_1^*
		+ 2\, s \left[\left(|A_0|^2 + 2 |A_{1}|^2\right) A_{-1} + A_0^2A_1^*\right] 
	\right\}
\end{aligned}
\label{eq:CMT}
\end{equation}
\end{widetext}
The dot denotes the derivative in $\xi$ and, stemming from $h[a]$, $s=\mathrm{sign}\,{\Omega}$. 
It is easy to verify that $U=|A_0|^2+|A_1|^2+|A_{-1}|^2$ is conserved ($\dot{U}=0$).  

\subsection{BFI sidebands}
In the  limit $|A_{\pm 1}|\ll |A_0|$ and $A_0(\xi) = \sqrt{U_0}\exp(-i U_0\xi)$, posing $A_{\pm 1}= u_{\pm 1} \exp(-i U_0\xi)$, we obtain the linearized system for $\mathbf{u}=[u_1, u_{-1}^*]^T$, $\dot{\mathbf{u}}=i \Sigma \mathbf{u}$, with
\begin{multline}
\Sigma\equiv\\
\begin{bmatrix}
	\frac{\Omega^2}{2}-U_0-(8-2s)\epsilon\Omega U_0 & -U_0 (1-2(s-\beta)\epsilon\Omega)\\
	U_0 (1-2(s+\beta)\epsilon\Omega) & 
	-\frac{\Omega^2}{2}+ U_0-(8+2s)\epsilon\Omega U_0
	\end{bmatrix},
	\label{eq:BFIODEs}
\end{multline}
which permits to recover the well-known dispersion relation ($\mathbf{u}\sim\exp(i\kappa \xi)$), at order $\epsilon$ 
\[
\kappa \approx -8\epsilon\Omega\eta_0 
\pm \frac{1}{2}|\Omega|\left[		
	\left(\Omega^2-4U_0(1-2s\epsilon\Omega)\right)
\right]^\frac{1}{2}
\]
Thus the peak linear gain $g_\mathrm{M}=U_0(1-2\epsilon\sqrt{2U_0})$ occurs for $\Omega_\mathrm{M}\approx\pm\sqrt{2U_0}(1-\frac{3}{2}\sqrt{2U_0}\epsilon)$ (i.e.~the value employed in the above simulations); we also recover that the instability bandwidth (where $\mathrm{Im}\,\kappa\neq 0$) shrinks to $[0,2\sqrt{U_0}(1-2\sqrt{U_0}\epsilon)]$ due to $h[a]$, as it is well-known since the original presentation of the Eq.~\eqref{eq:Dysthe1}  \cite{Dysthe1979,Lo1985}. 

An often disregarded aspect is the symmetry of sidebands. The eigenvector  of the matrix in Eq.~\eqref{eq:BFIODEs} corresponding to the faster-growing mode (i.e.~the eigenvector direction at $\Omega_\mathrm{M}$) is $|u_1/u_{-1}|_\mathrm{M}\approx1+2\beta\epsilon\sqrt{2U_0}$, so for $\beta=0$ the growth of BFI is symmetric. We thus can assert that  $g[a]$ is responsible for the asymmetry of the sideband growth, favoring the up-shifted mode. This is consistent with the nonconservation of $P$ mentioned above.

\subsection{Reduction to one degree of freedom}

We study the nonlinear stage of BFI, assuming now $U_0=1$. From the analysis presented above, the one unstable mode approximation corresponds to $\Omega \geq 1-2\epsilon$ (upper half of the instability bandwidth). 
By taking $A_m(\xi)=\sqrt{\eta_m(\xi)}\exp(i\phi_m(\xi)))$ ($m=0,\pm1$), we can show that  the system \eqref{eq:CMT} can be reduced to three variables, namely the sideband amplitude $\eta=\eta_1+\eta_{-1}=1-\eta_0$, the relative phase $\psi= (\phi_1+\phi_{-1}-2\phi_0)/2$ and the sideband unbalance $\alpha=\eta_1-\eta_{-1}$. If $\beta=0$, $\dot{\alpha}=0$ and Eq.~\eqref{eq:CMT} can be reduced to an integrable one-degree-of-freedom system in the canonical variables $(\eta,\psi)$  ($\dot{\eta}=\frac{\partial H}{\partial \psi}$, $\dot{\psi}=-\frac{\partial H}{\partial \eta}$), with Hamiltonian 
\begin{multline}
H(\psi,\eta) = -\left(\frac{\Omega^2}{2}  -\sigma  + \epsilon 4\Omega\alpha\right)\eta - \frac{3}{4}\eta^2 \\+  \epsilon  s\Omega\eta^2 + \sigma(1-\eta)\left[\eta^2-\alpha^2\right]^\frac{1}{2}\cos 2\psi
\label{eq:hamiltonian}	
\end{multline}
with $\sigma=(1-2s\epsilon\Omega)$. This integrable one degree-of-freedom reduction represents only a minor correction to the conventional NLS truncation  \cite{Trillo1991c}, where $\sigma=1$. Thus the phase space exhibits a heteroclinic structure; a separatrix connecting to hyperbolic fixed points implies the existence of two regimes: closed orbits inside, around a center fixed point, which correspond to period-one recurrence, and open orbits outside, which correspond to period-two (or phase-shifted) recurrence  \cite{Akhmediev1986,Akhmediev1987b,Kimmoun2016}, see also the insets in Fig.~\ref{fig:phaseplanecomparison} below.

The $\beta\neq 0$ case is ruled by
\begin{equation}
	\begin{aligned}
			\dot{\eta} &= -2 \sigma
		(1-\eta)\left[\eta^2-\alpha^2\right]^\frac{1}{2}\sin 2\psi \\
			\dot\psi &= \frac{\Omega^2}{2} - 1 + \frac{3}{2} \eta  + 2\epsilon\Omega\left(s(1-\eta)+\frac{4-\beta}{2}\alpha\right)\\
			&+ \sigma\left[\eta^2-\alpha^2\right]^\frac{1}{2}\cos2\psi\\
			&-\sigma\eta(1-\eta)\left[\eta^2-\alpha^2\right]^{-\frac{1}{2}}
			\cos 2\psi + \\
			&+2\beta\epsilon\Omega (1-\eta)\alpha\left[\eta^2-\alpha^2\right]^{-\frac{1}{2}}\cos 2\psi\\
			\dot{\alpha} &= -4\beta\epsilon \Omega(1-\eta)\left[\eta^2-\alpha^2\right]^\frac{1}{2}\sin 2\psi
		\end{aligned}
\label{eq:1Dnonintegrable}
\end{equation} 
The non-conservation of momentum in Eq.~\eqref{eq:Dysthe1}, see Eq. \eqref{eq:PoverN}, is reflected to a non-conserved unbalance of the sidebands and a modification of the phase dynamics. Both effects are perturbative, $\mathcal{O}(\epsilon)$, but have a measurable impact on the system evolution. Importantly we obtain a closed form for $\dot\alpha$, thus simplifying the qualitative and quantitative understanding of the evolution of HONLS and its further extensions.

In order to assess the deviation from integrability, we take $\eta(0)=0.01$, $\psi(0)=0$, $\alpha(0)=0$, $\beta=s=1$, $\epsilon=\sqrt{2}/20$ and compare the solution of Eq.~\eqref{eq:1Dnonintegrable} with the level sets of $H$, Fig.~\ref{fig:3wavephasespace}.  {For $\alpha=0$ in Eq.~\eqref{eq:hamiltonian}, the separatrix connects the two hyperbolic fixed points at $\pm \frac{1}{2}\cos^{-1}\left(\frac{\Omega^2}{2\sigma}-1\right)$, $\eta=0$; the center is at $\psi\equiv 0\pmod{\pi}$, $\eta_c= \frac{4\sigma - \Omega^2}{3+4\sigma-4\epsilon |\Omega|}$}. Albeit the non-conservation of $\alpha$ represents a sideband unbalance, it does not change the topology of the phase space (it is non-resonant).


\subsection{Comparison to simulations}

The relevance of the truncated model is further validated by comparing its solutions to the evolution of the sidebands according to the HONLSe. We use the same parameters as above, apart from the phase, for which we show the two different families of solutions (period-one for $\psi(0)=0$, period-two for $\psi(0)=\pi/2\mod{\pi}$). 
\begin{figure}[hbtp]
\centering
\includegraphics[width=0.5\textwidth]{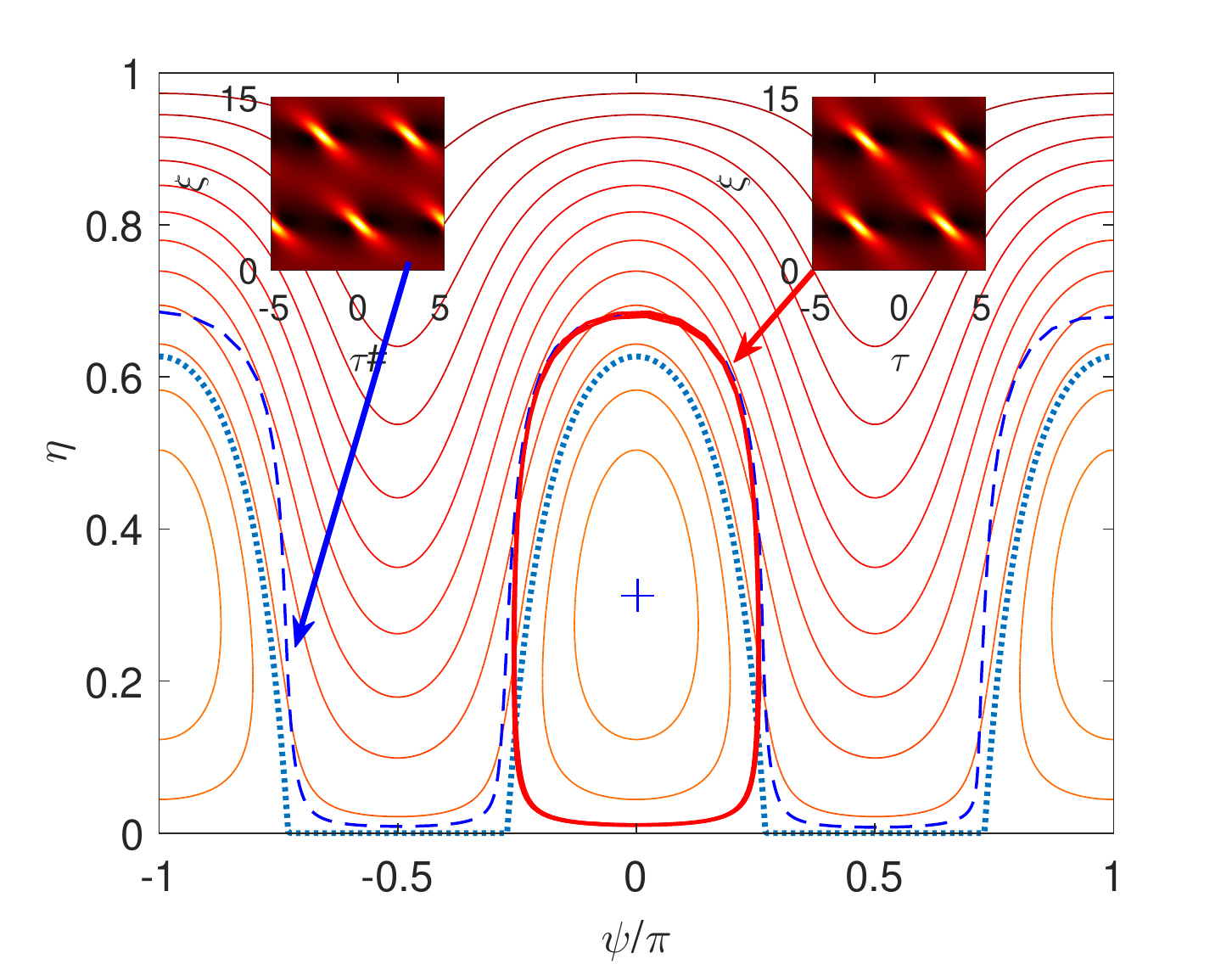}
\caption{Representation of solutions of HONLSe on the phase plane of the one degree-of-freedom model of Fig.~\ref{fig:3wavephasespace}. The red solid line correspond to the inside of the separatrix ($\psi(0)=0$, \emph{period-one}), the blue dashed line to the outside ($\psi(0)=\pi/2$, \emph{period-two}). In spite of a deviation from three-wave truncation, the topology is conserved. The two insets represent $|a|^2$ in the $(\tau,\xi)$ plane. As indicated by arrows, the ``phase-shifted'' solution (top-left) corresponds to the blue dashed curve on the phase space; the top-right  to the red solid curve, i.e.~non-phase shifted solution.}
\label{fig:phaseplanecomparison}	
\end{figure}
In Fig.~\ref{fig:phaseplanecomparison} we plot the HONLSe results in the three-wave phase-space.
The degree of superposition between the level sets of $H$ defined above and the HONLSe solutions is quantitatively good only for  conversion to higher-order sidebands smaller than 10$\%$, but the topology is completely preserved. As soon as a significant amount of energy is transferred to modes at $\pm 2\Omega$, the discrepancy grows. We verified numerically that the period of recurrence is under- or over-estimated by less that $10\%$ (around the maximum conversion frequency, not shown).

Larger values of steepness ($\epsilon>0.2$, not shown) severely perturbs the conservation of $P/N$ and, after 3 or more recurrence cycles, can sporadically break the regularity of the evolution and lead to the crossing of the separatrix. For those values, anyway, the steep pulses attained during the breathing cycle ($|a|\approx 3$) greatly enhance the probability of wave-breaking, so that the physics is beyond the validity range of Eq.~\eqref{eq:Dysthe1}. 

Then it is important to investigate the predictability of the spectral shift. 
\begin{figure}
 \centering
 \includegraphics[scale=0.5]{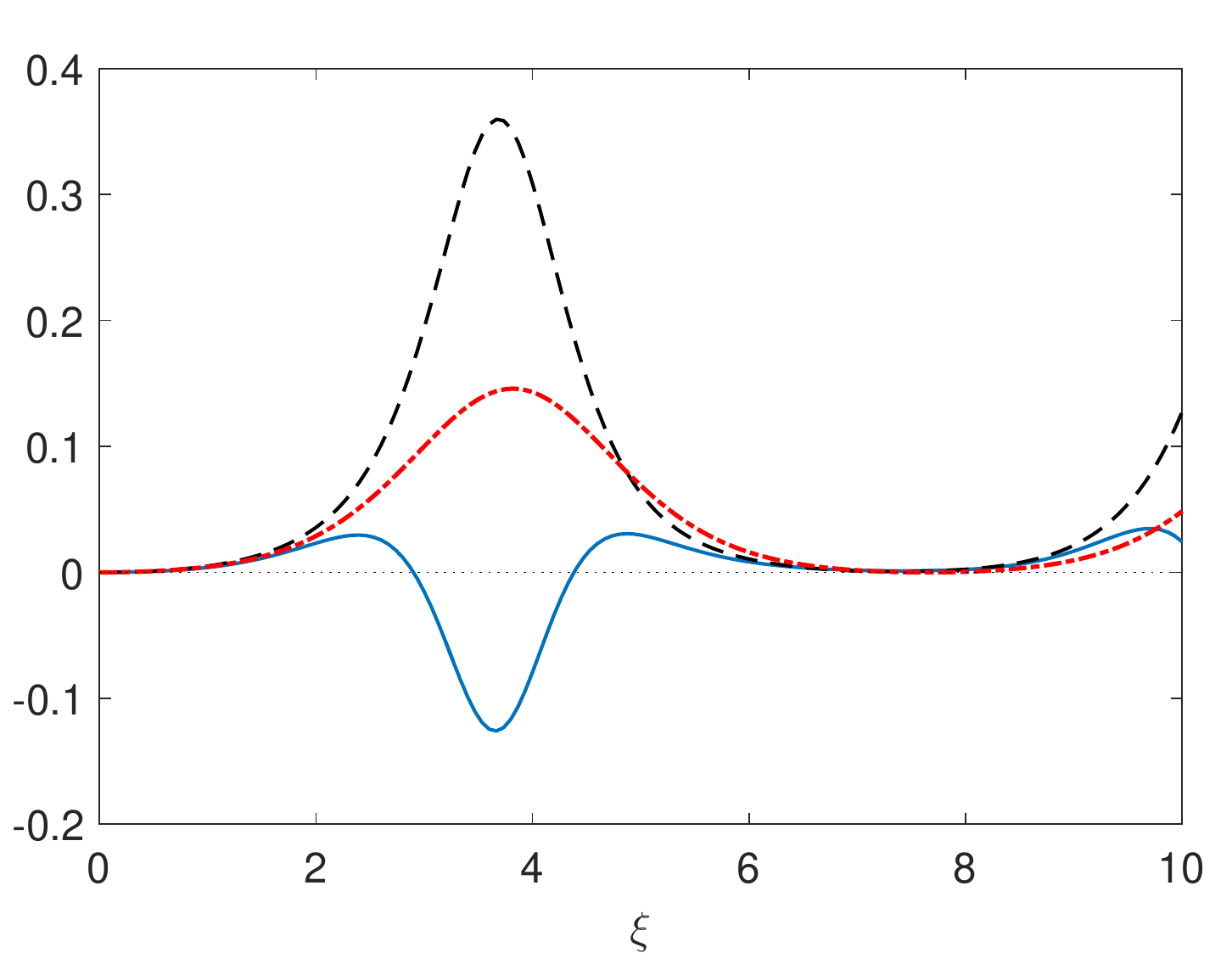}
 \caption{Comparison of the peak spectral shifts $\Pi$ and $\Pi_3$ obtained, respectively, from the full HONLS simulation (blue solid line) and the truncated three-wave model (red dash-dotted line). The spectral mean $P/N$ is also shown (black dashed line).  $\psi(0)=0$ while all other parameters are as above. 
}\label{fig:spectralshift}
 \end{figure}
{Let $\Pi_3\equiv\Omega\alpha$: this is a truncated version of $P/N$, because $N$ is constant and 
$P = \sum_{n=-\infty}^\infty {n\Omega |\hat{a}(n\Omega)|^2}$.
}
In Fig.~\ref{fig:spectralshift} we compare these two approximations of the spectral mean with the peak shift $\Pi$ extracted from simulations. As we showed in Fig.~\ref{fig:dysthe1}, the spectral peak exhibits a upshift/downshift transition. This is ascribed to the $f[a]$ contribution, which in turn enters in Eq.~\eqref{eq:1Dnonintegrable} as a pure phase shift, so it cannot provide a net effect on $\alpha$. Despite $\Pi_3$ resembles $\Pi$ in its definition, it describes instead the sideband unbalance caused by $g[a]$ in the three-mode truncation. The dash-dotted line actaully approximates the dashed curve of $P/N$. 

This consolidates the view that a definition of spectral shift cannot rely on the spectral peak position: this in turn  depends on the definition of the reference frame (a correction of group velocity) rather than on dynamical effects (the privileged direction of four-wave mixing, which on average points towards higher frequencies).

The correct definition of (here temporary) up(down)-shift must thus rely on the $P/N$ dynamics, which is notably appropriately described even by our low-dimensional approximation. The inclusion of more Fourier components (we verified it in the five-wave truncation) only slightly improves the approximation: the kinematic effect depends thus on a complex interplay of many stable and unstable modes, while the spectral shift is a global average property. As shown above, by comparing Figs.~\ref{fig:dysthe1} and \ref{fig:dysthe2}, for a smaller BFI gain the conversion to higher-order modes is smaller and the upshift is dominant, consistently with our analysis.

\section{Conclusions}

We analyzed a low-dimensional truncation of the so-called Dysthe equation in its two versions: the one which conserves only the norm (surface elevation HONLS) and the one which conserves also momentum and Hamiltonian (velocity potential HONLS). This latter naturally reduces to a one degree-of-freedom integrable Hamiltonian system, while the former includes a closed-form non-resonant perturbation which breaks the integrability, but not the heteroclinic behavior. 
This low dimensional (albeit non-integrable) system of ODEs permits to model correctly the \emph{temporary spectral upshift} in the spectral-mean sense of the nonlinear stage of BFI and to distinguish it from other kinematic corrections, which manifest themselves as a frequency downshift.

{In conventional wave-tanks, the surface elevation is the only accessible quantity, measured directly by mechanical gauges or detected by video-recording techniques. The modulated carrier wave is composed of free modes, whose envelope is ruled by the HONLSe, and bound modes, which oscillate at harmonic multiples of the first and follow them perfectly. The reconstruction of the envelope from experimental data is normally limited to the fundamental wave and a few of its sidebands. In \cite{Fedele2012}, a gauge transformation is presented which allows us to eliminate the $g[a]$ term with a modification of nonlinearity (a quintic term). This equation can be easily studied by means of our approach, but in that case we would lose the direct correspondence between the evolution of the sidebands in the truncated model and in the envelope of the surface elevation. 
}
{
For these reason we believe that the proposed low-dimensional approach is more physically transparent and allows a better understanding of the experimental data.}
Moreover it is general and can be applied to other corrections of the NLS and HONLS, such as linear or nonlinear gain or dissipation, in order to ease the interpretation of the complex experimental data collected in the hundred meter long wave-tanks which are being currently developed.

\begin{acknowledgements}
We acknowledge the financial support from the Swiss National
Science Foundation (Project No. 200021-155970) We would like to thank Debbie Eeltink and John D. Carter for fruitful discussions. 
\end{acknowledgements}

%

\end{document}